\documentclass[conference]{IEEEtran}
\usepackage{amsmath,amssymb,amsfonts}
\usepackage{hyperref}
\usepackage{graphicx}
\usepackage{xcolor}
\usepackage{booktabs}

\newcommand{\ra}[1]{\renewcommand{\arraystretch}{#1}}

\begin{document}

\title{Multi-car paint shop optimization with quantum annealing}
\author{\IEEEauthorblockN{Sheir Yarkoni\IEEEauthorrefmark{1}\IEEEauthorrefmark{2}, Alex Alekseyenko\IEEEauthorrefmark{3}, Michael Streif\IEEEauthorrefmark{1}\IEEEauthorrefmark{4}, David Von Dollen\IEEEauthorrefmark{3}\IEEEauthorrefmark{2}, Florian Neukart\IEEEauthorrefmark{1}\IEEEauthorrefmark{2}, Thomas B\"ack\IEEEauthorrefmark{2}}\\

\IEEEauthorblockA{\IEEEauthorrefmark{1}Volkswagen Data:Lab\\Munich, Germany}\\

\IEEEauthorblockA{\IEEEauthorrefmark{2}LIACS, Leiden University\\Leiden, The Netherlands}\\

\IEEEauthorblockA{\IEEEauthorrefmark{3}Volkswagen Group of America\\San Francisco, CA, USA}\\

\IEEEauthorblockA{\IEEEauthorrefmark{4}Institute of Theoretical Physics, University Erlangen-N\"urnberg \\N\"urnberg, Germany}

}







\newcommand{\sheir}[1]{\textcolor{blue}{\textbf{Sheir}: #1}}

\maketitle

\begin{abstract}
    We present a generalization of the binary paint shop problem (BPSP) to tackle an automotive industry application, the multi-car paint shop (MCPS) problem. The objective of the optimization is to minimize the number of color switches between cars in a paint shop queue during manufacturing, a known NP-hard problem. We distinguish between different sub-classes of paint shop problems, and show how to formulate the basic MCPS problem as an Ising model. The problem instances used in this study are generated using real-world data from a factory in Wolfsburg, Germany. We compare the performance of the D-Wave 2000Q and Advantage quantum processors to other classical solvers and a hybrid quantum-classical algorithm offered by D-Wave Systems. We observe that the quantum processors are well-suited for smaller problems, and the hybrid algorithm for intermediate sizes. However, we find that the performance of these algorithms quickly approaches that of a simple greedy algorithm in the large size limit. 
     
\end{abstract}

\begin{IEEEkeywords}
quantum annealing, quantum computing, optimization, sequencing
\end{IEEEkeywords}

\section{Introduction}
Quantum computing has the potential to significantly impact both research and industry in a variety of disciplines. Quantum algorithms developed expressly for fully error-corrected quantum computers are among the most promising, specifically Shor's algorithm for factoring~\cite{shor} and Grover's algorithm for unstructured search~\cite{grover}, which are both asymptotically faster than their classical counterparts. Without error correcting schemes, the advantages of quantum computing are more difficult to discern. Recently, a significant milestone of quantum supremacy was reached by Google's quantum computer, when it was shown to sample from random quantum circuits faster than any classical algorithm~\cite{quantumsupremacy}. In the near term, specific focus has been dedicated to developing useful algorithms for so-called Noisy Intermediate-Scale Quantum (NISQ~\cite{nisq}) processors. Various companies such as Google, IBM, and others have built gate-based NISQ computers with up to dozens of qubits~\cite{quantumsupremacy, ibm_qpu}. Companies such as D-Wave Systems have adopted an alternative approach, pursuing instead the physical implementation of quantum annealing algorithms, an adaptation of Adiabatic Quantum Optimization. Originally proposed as a heuristic optimization algorithm~\cite{nishimori}, the field of quantum annealing has grown significantly over time, with many potential applications being showcased in fields such as quantum simulation~\cite{ktlattice, phasetransitions, vwchemistry}, quantum machine learning~\cite{quantum_clust, qbm, timeseries}, optimization~\cite{jobshop, trafficflow, flightscheduling}, and more~\cite{portfolio, advertising}. The majority of both gate-model and quantum annealing applications currently involve solving an optimization problem, typically formulated as either an Ising Model using spin variables $s \in \{-1, 1\}$, or a Quadratic Unconstrained Binary Optimization (QUBO) problem with binary variables $x \in \{0, 1\}$. These two models are equivalent under a simple change of basis, and are known to be NP-hard to minimize in the worst case~\cite{barahona}. As such, many interesting and complex optimization problems can be posed in one of these models~\cite{Lucas}. Ising Models are formulated as a system of interacting spins:
\begin{equation}
    H(s) = \sum_i h_i s_i + \sum_{i<j} J_{ij} s_is_j,
\end{equation}
where $h_i$ and $J_{ij}$ are real-valued numbers defining the strength of interactions between neighboring spins $s_i, s_j$. For QUBOs, we define:
\begin{equation}
    \mathrm{Obj}(x, Q) = x^\mathrm{T} \cdot Q \cdot x, 
\end{equation}
where $x$ is a vector of binary variables, and here $Q$ is an upper-triangular matrix defining the relationships between them. The initial work of constructing an optimization problem suitable for quantum computers is finding a valid representation in either Ising or QUBO form. In quantum annealers, these optimization problems are encoded by programming sets of qubits and couplers. The quantum processor then initializes the system of qubits in an easy-to-prepare ground state, and slowly transitions the Hamiltonian to represent the optimization problem as specified by the user. The physical and mathematical background motivating the algorithm is beyond the scope of this work, and we refer the reader to~\cite{manufacturedspins}.  

Despite the proof-of-concept projects presented above which have been tested on quantum processors, a large-scale application has yet to be shown. Partly, this can be attributed to the mismatch between the core optimization problem and the limited scope of quantum optimization algorithms. Many industry problems involve a mix of integer, discrete, and continuous variables and constraints, while NISQ processors solve small-scale binary optimization problems. Therefore, it is important to find candidate problems that not only benefit from quantum computing theoretically, but also are amenable to the technology practically. We therefore present an automotive industry problem, the multi-car paint shop (MCPS) problem. This optimization problem is an extension of the binary paint shop problem (BPSP) which has been studied in the context of other quantum algorithms~\cite{bpsp}. The MCPS problem represents the real-world version of the BPSP, and is a crucial step in the process of car manufacturing at Volkswagen. In this study we only solve problem instances obtained from a factory in Wolfsburg, Germany. We employ a variety of classical solvers as competition benchmarks, and compare the results to D-Wave quantum processing units (QPUs) and a hybrid quantum-classical algorithm.  \\
The rest of the paper is structured as follows: in Section~\ref{sec:previous} we review the relevant existing literature for the paint shop problem. Section~\ref{sec:mcps} motivates and defines the multi-car paint shop problem, and derives the Ising model for the problem. In Section~\ref{sec:experiments} we outline the experimental setup and data preparation, and Section~\ref{sec:discussions} presents the results of our analysis. Lastly, Section~\ref{sec:conclusions} discusses the conclusions of the scientific analysis and the lessons learned from this study.  

\section{Previous works}
\label{sec:previous}
The paint shop problem refers to a set of combinatorial optimization problems in the automotive industry. The objective is to color a given sequence of cars with a fixed number of colors such that the total number of color switches is minimized. This simple problem poses interesting scientific complexity-theoretic questions, which in turn have real impact for solving such problems in practice. The paint shop problem was originally posed by Epping et. al~\cite{paintshop} as a form of coloring problem. Some clarification on nomenclature: the given car sequence can also be referred to as a word, where each car is denoted by a character. In~\cite{paintshop}, it was shown that the paint shop problem is NP-complete in both the number of colors and cars in the sequence. Furthermore, results show that, for bounded numbers of colors and unique cars, there exists a polynomial-time dynamic programming solution to these instances. Subsequent work~\cite{paintshop2} extended these results, proving that even the simplest coloring version, with only two colors, is both NP-complete and APX-hard. Additional results show that a subset of problems meeting specific conditions can be solved in polynomial time.

Streif et. al~\cite{bpsp} investigated solving the restricted problem, where each ensemble of cars is exactly one pair and only two colors are used (binary paint shop problem, BPSP), using a quantum algorithm. Specifically it was shown that a Quantum Approximate Optimization Algorithm (QAOA) of fixed depth can outperform classical heuristics for solving the BPSP in the infinite size limit. The authors investigate multiple classical heuristics for solving the BPSP and give bounds on the asymptotic performance of each to show this result.

\section{Multi-car paint shop optimization}
\label{sec:mcps}
One of the steps in car production at Volkswagen is painting the car body before assembly. In general, this can be viewed as a queue of car bodies that enter the paint shop, undergo the painting procedure, and exit the paint shop. It is important to note that the area of the factory immediately proceeding the paint shop is typically assembly, where car components are assembled into the car bodies. Because of the many different models and configurations being produced, designing a sequence of cars to be assembled that is optimal (also known as the car sequencing problem) is a known NP-hard problem in itself~\cite{car_sequencing}. In practice it is imperative to solve the sequencing problem because of worker safety and regulatory issues, and we therefore treat the sequence of cars entering the paint shop as a fixed queue. However, the \emph{colors} assigned to the cars within a given sequence are still randomly distributed, and thus we can focus on optimizing them.

\begin{figure}[h]
    \centering
    \includegraphics[width=\linewidth]{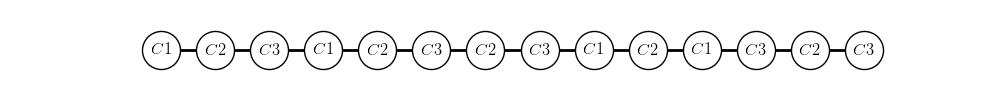}
    \includegraphics[width=\linewidth]{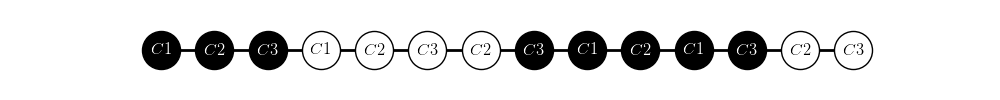}
    \includegraphics[width=\linewidth]{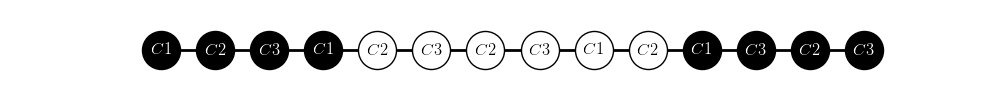}
    \caption{Simple example of a multi-car paint shop problem with three car ensembles $(C_1, C_2, C_3)$. The three corresponding orders are $k(C_1)=3, k(C_2)=2,$ and $k(C_3)=3$. \textbf{Top:} The fixed sequence of cars in the paint shop queue. \textbf{Middle:} Sub-optimal solution to the problem with 3 color switches. \textbf{Bottom:} Optimal solution to the problem with 2 color switches.}
    \label{fig:MCPS_figure}
\end{figure}

Each car body entering the paint shop is painted independently in two steps: the first layer is called the filler, which covers the car body with an initial coat of paint, and the final color layer is the base coat, which is painted on top of the filler. The base coat is the color that matches the final customer order: blue, green, etc. However, the filler has only two possible colors: white for the lighter base coats colors, and black for darker colors. We define a customer order as the number of cars of each configuration to be painted one of the color choices. We associate each coating step (filler or base coat) with a unique class of paint shop problems, each of which are NP-complete~\cite{paintshop, paintshop2}. The filler optimization is referred to as the \emph{multi-car paint shop} problem, where the cardinality of each set of unique cars in the sequence is unconstrained, but the cardinality of the color set is restricted to two (e.g., black or white). The base coat optimization is therefore an extension of the MCPS problem, where the cardinality of the color set is also unconstrained-- we call this version of the problem the \emph{multi-car multi-color paint shop} problem. In our work we focus on the filler optimization, the MCPS problem, which can be formulated natively as a binary optimization problem. We formally define the problem as follows:

\begin{table}[h!]
    \begin{tabular}{ll}
         Given: & a word $w$ defining the fixed sequence of $N$ cars ($w_i$ denotes \\
          & the $i$th character in $w$),  \\
          & \\
          & set of $C = \{C_1, \ldots, C_M\}$ unique car ensembles, \\
          & \\
          & binary choice of colors $\{W, B\}$, \\
          & \\
          & function $k(C_i)$ which defines the number of $C_i$ to be pain- \\
          & ted $B$ in $w$,\\
          & \\
          & function $f(w)$ to count the number of color switches in $w$, \\
          & \\
         such that: & $\#C_i|_{B} = k(C_i),~\forall C_i \in C$,\\
         minimize: & $f(w).$
         
    \end{tabular}
    \label{tab:mcps_def}
\end{table}
\newpage
In practice (i.e., in the paint shop), the information required to formulate this optimization problem is always available, as it is a necessary part of fulfilling customer orders. Therefore, this MCPS problem representation above corresponds exactly to the industrial use-case of paint shop optimization on the filler line. In Fig.~\ref{fig:MCPS_figure} we show a simple example of the MCPS problem with three car ensembles. 

Despite the focus of our work on the filler, we briefly address the optimization of the base coat. Although the filler and base coat are painted independently in separate locations, computationally the two problems are not separable. Abstractly, optimizing the base coat line (i.e., solving the multi-car multi-color paint shop problem) is a straightforward generalization of the MCPS problem: we can extend the binary color variables to discrete color variables, where $k(C_i)$ denotes the number of times each color appears in a car ensemble. In practice, it is useful to consider solving the multi-color problem \emph{after} solving the MCPS problem. Due to the aforementioned one-to-one mapping between base coat and filler color, it is still possible to permute the order of base coat colors within a contiguous sequence of filler colors to reduce the base coat color switches. Although discrete optimization problems can be represented as binary optimization problems, we do not solve the multi-color version of the problem and leave this work for future studies.    

\subsection{Ising model representation}
In principle, the formulation of the MCPS as an Ising model is straightforward: we start by representing every car in the sequence $w~(w_i)$ with a single spin variable ($s_i$). The spin up state denotes if the car is painted black, and the spin down state denotes if it is white. The Ising model which represents our problem can be divided into a hard constraint component and an optimization component. The optimization component is a simple Ising ferromagnet with $J=-1$ couplings between adjacent cars in the sequence:
\begin{equation}
H_A = -\sum_{i=0}^{N-2} s_is_{i+1}.
\end{equation}
This incentivizes adjacent cars to have the same color. The second component of our Ising model is the hard constraint, ensuring that the correct number of cars are colored white/black per customer orders. This is encoded in a second energy function, as a sum over independent $k$-hot constraint for each ensemble of cars $C_i$:
\begin{equation}
    H_B(C_i) = \left(\#C_i-2k(C_i)\right)\sum_i s_i + \sum_{i<j} s_is_j.
\end{equation}
Therefore, the final Ising model is the sum of the two components, $H_\mathrm{MCPS}~=~H_A~+~\lambda H_B$:
\begin{multline}
    H_\mathrm{MCPS} = -\sum_{i=0}^{N-2}s_is_{i+1} \\ + \lambda \sum_{C_i \in C} \left[\left(\#C_i-2k(C_i)\right)\sum_i s_i + \sum_{i<j} s_is_j\right],
    \label{eq:mcps_ising}
\end{multline}
with terms as previously defined in the MCPS problem statement. In order to ensure only valid configurations of spins are encoded in the ground state, it is necessary to scale $H_B$ by the factor $\lambda$~\cite{Lucas}. This enforces that it is never energetically favorable to violate a constraint to reduce the energy of the system. In our study we set $\lambda = N$, the total number of cars in our sequence, which guarantees this condition.
\section{Experimental methods}
\label{sec:experiments}
\subsection{Data sources}
The MCPS problem instances we used were generated from real data taken from a Volkswagen paint shop in Wolfsburg, Germany. The reason for this is two-fold: firstly, the main goal of this work is to test the viability of quantum annealing methods in solving industrial optimization problems. It is our goal to accurately capture the complexity of the industrial use-case without relying on simplifications or randomly-generated problem instances. Secondly, the paint shop currently operates on a first-come-first-served basis, where customer orders are entered into the queue as soon as they arrive. This guarantees a certain amount of randomness (although not uniformity) in the problem instances we solve: different car models, configurations, and base coat colors appear throughout the sequences in ways we do not control. Therefore, these conditions are suitable for our analysis. 

A total of 104,334 cars were used in this study. To reproduce real-world conditions as faithfully as possible, the car sequences used are multiple independent sets of car sequences, each representing one week of continuous production, which are stitched together as one continuous block. The data is collected over a period of one year, roughly once every six weeks, to avoid seasonal biases in customer order preferences. There are 121 unique car configurations in the data set. Of those, 13 of the configurations appeared only once in the data set, and therefore do not need optimization at all. Typically this indicates that either a custom configuration was built that cannot normally be ordered, or a prototype assembled for testing purposes. Rather than exclude these from the study, we include them in the optimization, because fixing a color at one location in the sequence influences the adjacent cars (and consequently the total number of color switches) in a non-trivial way. We do, however, eliminate the spin variable from the Ising model by conditioning on the color. 



\subsection{MCPS problem sizes}
To generate a variety of input sizes from the full data set, we partition the data into different sized sequences without permuting the car order. This is motivated by the amount of cars that need to be optimized for different purposes. For example, the paint shop used as a basis for this analysis has a queue capacity of roughly 300 cars. This is not the \emph{total} capacity of the paint shop, but rather the maximum number of cars that can be physically inside the paint shop queue before they are painted. We consider this a rough lower bound on the problem size for industrially-relevant problem instances. An upper bound is more difficult to establish. From a theoretical point of view, there is merit in investigating the behavior of quantum systems in the infinite size limit, as in~\cite{bpsp}. From an industrial perspective, car orders can be placed weeks to months in advance, which would yield problem sizes of $10^3-10^5$ variables. In reality, real-time and last-minute adjustments (due to manufacturing problems, supply chain issues, or imperfections in painting) can happen on a daily basis. We limit the analysis to problems of up to 3000 variables, which roughly corresponds to a few days worth of production. The data partitioning is performed by dividing the entire data set into equal chunks for each problem size $N$. Each partition is considered a candidate instance, yielding a total of $\lfloor \frac{104,344}{N_\mathrm{cars}}\rfloor$ partitions per problem size. We then mine these partitions and select suitable partitions to generate problem instances from. This is due to the fact that there could be little frustration within some partitions. For example, it is possible that all cars of any one configuration ($C_i$) \emph{all} need to be painted either black or white. While in general this is an accurate reflection of production, for experimental analyses this scenario is not useful. Therefore we deem a data partition to be a usable MCPS instance if the total number of non-fixed cars is at least $70\%$ of the cars in the partition. Meaning that, in a 10-car data partition, at least 7 of the cars must have the freedom of being painted either color. We show the total number of partitions for the various problem sizes and how many partitions were valid problem instances in Table~\ref{tab:problem_sizes}. For our experiments we randomly selected 50 valid instances to test at each $N$, except for the largest problem size of which we use all 34 valid instances.

\begin{table}[h]
    \centering
    \caption{Problem sizes and number of problem instances generated from the data set partitions.}
    \begin{tabular}{|c|c|c|}
        \hline
         \bf Problem size & \bf Num. partitions & \bf Num. instances \\ 
           \bf (cars) & & \bf ($\%$ of partitions) \\ \hline \hline
         10 & 10,433 & 172 $(1.6\%)$ \\ \hline
         30 & 3,477 & 418 $(12.0\%)$ \\ \hline
         100 & 1,043 & 756 $(72.5\%)$ \\ \hline
         300 & 347 & 341 $(98.3\%)$ \\ \hline
         1000 & 104 & 102 $(98.1\%)$ \\ \hline
         3000 & 34 & 34 $(100\%)$ \\ \hline
    \end{tabular}

    \label{tab:problem_sizes}
\end{table}

\subsection{Classical, quantum, and hybrid solvers}
We now review the solvers used in our experiments: \\
\textbf{Random.} Without optimization, we consider any assignment of colors to cars in a sequence where the orders are fulfilled to be a valid, but not necessarily optimal, solution to the problem. Thus, we can trivially generate random sets of valid solutions by uniformly assigning the color black to $k(C_i)$ cars for each car ensemble $C_i$. While far from optimal, this solution to the MCPS problem may be preferable if other steps in the car manufacturing process are valued over the painting step. This was indeed the case for the data obtained from the paint shop in Wolfsburg, and thus random valid solutions serves as the baseline the competition algorithms are tasked with beating. For our study, we generate $2N_\mathrm{cars}$ random valid solutions at each problem size to estimate the number of color switches that would occur naturally in the paint shop.\\  
\textbf{Black-first.} This is the simplest algorithm that is used to solve the MCPS problem. Starting at the beginning of the sequence, we greedily assign the color black to every car until a white color \emph{must} be assigned to the next car, meaning all orders $k(C_i)$ have been fulfilled. In greedily obtained solutions the number of color switches grows linearly with the number of cars and is sub-optimal except for a minority of cases~\cite{paintshop2, bpsp}. Nonetheless, it serves as a good benchmark for more sophisticated optimization algorithms, as the improvement over random grows with the problem size.\\
\textbf{Simulated annealing.} The simulated annealing (SA) metaheuristic is a well-known optimization routine inspired by the physical annealing of metals~\cite{sa}. Starting from an initial random state, potential solutions' individual variables are flipped randomly according to an acceptance criterion at every step (called a sweep) based on a ``temperature'' parameter. As this temperature is ``cooled'', it becomes increasingly unlikely that an energetically unfavorable move is accepted, at which point the search is terminated. This algorithm has been used in previous quantum computing benchmarking studies~\cite{ttt, goodbadugly, localruggedness}. We use the open-source D-Wave Python library Dimod implementation of SA~\cite{dimod}. \\
\textbf{Tabu search.} Another metaheuristic used to solve QUBOs based on individual variable flips is called Tabu search~\cite{tabu1, tabu2}. In each candidate solution, variables' states are flipped based on the likelihood they contribute to the optimality of the solution, and revisiting the same variables is discouraged, based on the ``tabu tenure'' list. Flips which worsen the solution quality are admissible if no other move is possible, which provides a trade-off between quality of solution and globality of the search. The Dimod Python library contains an implementation of Tabu search as well~\cite{dimod}. \\
\textbf{D-Wave 2000Q.} D-Wave QPUs have been tested extensively in literature by embedding Ising models directly onto the processors using an approach of minor-embedding. This process maps logical variables in the Ising model to chains of qubits in the target graph. The maximum size of problems which can then be solved directly by the QPU is limited by both the number of qubits in the processor and their connectivity. The D-Wave 2000Q generation of QPUs have a relatively sparse connectivity graph, named Chimera, where each qubit has at most degree 6. The processor used in this study had 2041 functional qubits. \\
\textbf{D-Wave Advantage.} The newest-generation QPU provided by D-Wave has a different topology and a significantly higher qubit count than its predecessor. The new topology, named Pegasus, has a maximum degree of 15, and the QPU used in our study contained 5436 functional qubits. For further information regarding the D-Wave QPU topologies and the differences between them, we refer the reader to~\cite{pegasus}.\\
\textbf{D-Wave Hybrid Solver.} At large instance sizes (300 cars and higher) it was no longer possible to embed problems directly onto both D-Wave QPUs. Therefore we employed a state-of-the-art hybrid quantum-classical algorithm developed and maintained by D-Wave System called the Hybrid Solver Service (HSS). This algorithm is designed to solve arbitrarily-structured QUBOs and Ising models of up to $10^4$ variables, while also leveraging access to a QPU in its inner loop. It is accessible via the same API as D-Wave's QPUs. Despite the power of such a hybrid algorithm, the QPU it uses cannot be programmed directly by the user, and thus we treat this algorithm as an optimization black-box with a single timeout parameter. 

\section{Results and discussion}
\label{sec:discussions}

\begin{figure}[h]
    \centering
    \includegraphics[width=\linewidth]{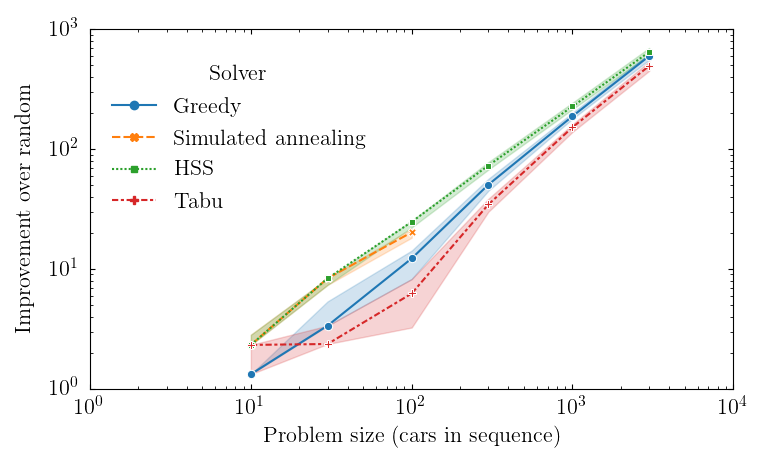}
    \caption{Number of color switches within the sequence shown as improvement over random configurations of orders.}
    \label{fig:results}
\end{figure}

We interpret our results relative to two different regimes: small-scale (10-100 cars) and industrial (300-3000 cars). Each solver used in these experiments was tuned in good-faith, but not necessarily optimally. Meaning, considerable effort was made to ensure solvers were being used to their strengths, but fully optimizing over all sets of hyperparameters for the solvers was deemed out of scope. We compared all solvers' performance in terms of their ``improvement'' over the random solver, defined as the difference in $f(w)$ between the best solution obtained by each solver and the random solver. This metric is representative of the real-world expected improvement using each of the solvers. In Table.~\ref{tab:results_solvers} we report the median for each solver. We quote the median to be less susceptible to tails of the distribution. The results therefore represent the typical MCPS case at each problem size, rather than the expected value of each solver's overlap with the ground states. The solutions obtained from all solvers were post-processed (if needed) to ensure that the $k(C_i)$ constraint was satisfied in each problem. We show how often this occurred in Table~\ref{tab:results_solvers} as well. We consider this necessary in order to interpret the solutions relative to the application, since it is trivial to reduce the number of color switches by ignoring the customer order constraint. In Fig.~\ref{fig:results} we highlight the empirical scaling of our results in the industrial limit.  

Solving problems directly with both QPUs was only possible for problem sizes 10-100. Embeddings were generated using the standard D-Wave embedding tool Python package~\cite{minorminer}. The chain strengths required by each QPU was different depending on the length of the chains in the embeddings. We calculated the algebraic chain strength $\mathrm{chain}_i = |h_i| + \sum_{j \in \mathrm{adj}(i)}|J_{ij}|$ for every spin in the Ising model, and introduce an additional scaling parameter $s = [0.1, 0.2, \ldots, 1]$. Thus, the chain strength is defined as $s\cdot \mathrm{max(chain}_i)$, where $s$ was optimized per problem size using a subset of the instances (10 per size), and $50\cdot N$ samples per instance. Optimal $s$ was defined as the value which yielded the highest frequency of valid solutions relative to the constraints of the MCPS problem in Eq.~\ref{eq:mcps_ising} (not chain breaks). We found that the D-Wave Advantage QPU had optimal $s=0.3$, whereas the D-Wave 2000Q QPU had optimal $s=0.45$. This is consistent with the fact that the Advantage QPU required shorter chains to embed the same Ising models as the 2000Q. Each QPU was then sampled for $500\cdot N$ samples per problem size $N$, with annealing time $t_a = 1\mu s$. We used $N$ spin-reversal transforms for each sample set, as it has been shown that there are diminishing returns between 100-1000 samples per transform~\cite{temperature}.

We found that for the 10 car instances both QPUs matched the consensus best results between all solvers (median of $f(w) = 2$), and for the 30 car instances very near the best results ($f(w)=5$ as opposed to SA's and HSS's 4). For the 100 car instances, with 50,000 samples per problem, the 2000Q QPU found valid solutions for 22/50 instances, as opposed to 37/50 for the Advantage QPU. From this we conclude that 50,000 samples is insufficient for the QPUs, but due to limited time availability we could not take more samples. We note that it was also possible to embed 47/50 of the 300 car instances onto the Advantage QPU. However, due to the poor performance on smaller sizes with limited resources, we did not evaluate those problems. Furthermore, due to the limited problem sizes that could be solved with the QPUs and the quality of results, we do not include these results in Fig.~\ref{fig:results}. \\
The Tabu solver was given $\lfloor N/3\rfloor$ seconds per problem as its timeout parameter, and all other parameters were set to their default value. This solver struggled to find valid solutions past the 10 car instances. Due to the post-processing technique, which greedily corrected each sample to satisfy order $k(C_i)$, the Tabu results were essentially a worse version of the greedy algorithm at all problem sizes but the smallest. \\
Simulated annealing (SA) has many tunable hyperparameters: number of sweeps ($N_\mathrm{sweeps}$), number of samples ($N_\mathrm{samples}$), and (inverse) temperature ($\beta$) schedule. In our experiments we fixed the schedule to $\beta = [0.01, 10]$, interpolated geometrically using $N_\mathrm{sweeps}$. We set $N_\mathrm{sweeps} = 10\cdot N$ and $N_\mathrm{samples} = 20\cdot N$, where $N$ is the number of cars. The solutions obtained by SA (shown in Fig.~\ref{fig:results}) were consistently better than the greedy algorithm. Using the given parameters SA was able to provide valid solutions for 50/50, 49/50, and 44/50 for the 30, 50, and 100 car instances. However, the timescales necessary to obtain results were prohibitive from extending the experiments: 300 variable problems were terminated after running for 24 hours without returning a solution. We include the SA results in Fig.~\ref{fig:results} due to their high quality at small sizes. Using a single-threaded SA implementation, run-time of the algorithm was on the order of seconds to minutes for the 10 and 30 car instances, and between 1-3 hours for the 100 car instances. We note that SA was the only solver which was allotted quadratically scaling computing resources: both sweeps and samples scaled with $N$. This is necessary for SA to be competitive, and exemplifies the trade-off between results quality and algorithmic run-time when using heuristics.  \\ 
The D-Wave HSS was given equal time to the Tabu solver, given its only parameter is the timeout: $\lfloor N/3\rfloor$ seconds per problem. The HSS was the only solver to consistently provide better solutions than the greedy algorithm for all problem sizes. The improvement continued to grow with increasing problem size, shown in Fig.~\ref{fig:results}. However, the gap between the HSS and the greedy algorithm shrank with increasing problem size, performing only slightly better than greedy for the 3000 car instances.

Despite the simplicity of the Ising model representation of the MCPS problem, almost all solvers exhibited difficulty in finding valid solutions. This is particularly evident in the performance of the two QPU models. The results degrading rapidly from 30 to 100 cars indicate that the problem became more difficult to solve disproportionately to the increase in system size. The SA and Tabu solvers exhibited similar trends, despite the increase in resources allotted to them. While the HSS was the best-performing algorithm, it also missed valid solutions for some problems of intermediate size (100 and 300 car instances). We identify two possible issues: first is the connectivity of the problem graph. Each order $k(C_i)$ requires a separate $k$-hot constraint which is represented using a fully-connected graph. This yields sub-cliques within each problem that increase with the problem size. In QPUs, denser problems create longer chains and higher chain strengths. For classical solvers, these sub-cliques create rugged landscapes which make single-flip optimization algorithms significantly less useful. Therefore, that the maximum sub-clique of the MCPS problem graph increases as a function of the number of cars is a bottleneck for performance.
Secondly, the normalization terms necessary to encode the MCPS problem as an Ising model also scale with problem size. In Eq.~\ref{eq:mcps_ising}, we set $\lambda=N$ to ensure the constraints are valid in the ground state of the Ising problem. Therefore, we observe that the numerical precision necessary to formulate the MCPS problem scales as $1/N$. This effect compresses the gaps between the local (and global) minima, making it harder to differentiate between them. Using a direct embedding approach for QPUs requires even higher precision due to chains: the chain strengths scale with $N$, and therefore the encoding precision as $1/N^2$. This may be prohibitively low in large-scale problems and analog devices.

\begin{table*}[t]
\scriptsize
\centering
\caption{Results for all solvers. We present the percentage of problems for which each solver found valid solutions ($\%_\mathrm{valid}$) and the median number of color switches $\left(f(w)\right)$ for each problem size $N$ cars. \label{tab:results_solvers}}

\ra{1}
\begin{tabular}{@{}rrrrrrrrrrrrrrr@{}}\toprule

& \multicolumn{2}{c}{\bf Greedy} & \multicolumn{2}{c}{\bf HSS} & \multicolumn{2}{c}{\bf Tabu} & \multicolumn{2}{c}{\bf SA} & \multicolumn{2}{c}{\bf 2000Q} & \multicolumn{2}{c}{\bf Advantage} & \multicolumn{2}{c}{\bf Random} \\




\cmidrule{2-3} \cmidrule{4-5} \cmidrule{6-7} \cmidrule{8-9} \cmidrule{10-11} \cmidrule{12-13} \cmidrule{14-15} & $\%_\mathrm{valid}$ & $f(w)$ & $\%_\mathrm{valid}$ & $f(w)$ & $\%_\mathrm{valid}$ & $f(w)$ & $\%_\mathrm{valid}$ & $f(w)$ & $\%_\mathrm{valid}$ & $f(w)$ & $\%_\mathrm{valid}$ & $f(w)$ & $\%_\mathrm{valid}$ & $f(w)$ \\ \midrule

 $N = 10$ \\
  & $100\%$ & 3 & $100\%$ & 2 & $86\%$ & 2 & $100\%$ & 2 & $100\%$ & 2 & $100\%$ & 2 & $100\%$ & 4.325 \\

 $N = 30$ \\
  & $100\%$ & 9 & $100\%$ & 4 & $26\%$ & 10 & $98\%$ & 4 & $100\%$ & 5 & $100\%$ & 5 & $100\%$ & 12.367 \\

 $N = 100$ \\
  & $100\%$ & 23 & $92\%$ & 10.5 & $0\%$ & 29 & $88\%$ & 15 & $44\%$ & 31 & $74\%$ & 28.5 & $100\%$ & 35.25 \\

 $N = 300$ \\
  & $100\%$ & 57 & $96\%$ & 34.5 & $0\%$ & 73 & - & - & - & - & - & - & $100\%$ & 107.58 \\

 $N = 1000$ \\
  & $100\%$ & 160 & $100\%$ & 121 & $0\%$ & 196 & - & - & - & - & - & - & $100\%$ & 348.35 \\

 $N = 3000$ \\
  & $100\%$ & 455 & $100\%$ & 406.5 & $0\%$ & 558.5 & - & - & - & - & - & - & $100\%$ & 1054.41 \\








\bottomrule
\end{tabular}
\end{table*}


\section{Conclusions}
\label{sec:conclusions}
In this paper we presented a potential application for quantum computing, the multi-car paint shop problem. We used real-world data from a paint shop in Wolfsburg, Germany, to generate problem instances and benchmark the performance of competition QUBO and Ising solvers. We found that for small problems of up to 30 cars the two QPUs tested, the hybrid algorithm, and simulated annealing were able to significantly improve results over production conditions (random distributions of colors). At intermediate and large problem sizes (100 cars and up), only the D-Wave HSS was able to consistently beat the greedy algorithm. Although the improvement over random grew with system size, the performance of the HSS and the greedy algorithm converged in the large size limit. We find that our methods demonstrated the potential of quantum annealing, and specifically the viability of hybrid quantum-classical approaches in solving industrially relevant optimization problems. We identified potential bottlenecks in performance and gave insights into how the formulation of the MCPS problem-- both in terms of problem graph structure and precision-- could affect the performance of the different solvers we used in our experiments. We note that in our experimental setup, we only tested whether the presented algorithms were capable of solving the industrial problems we generated, and make no claims in regards to the ultimate scaling of these algorithms. We defer a more thorough evaluation of algorithmic scaling and performance optimality for future work.\\
Overall, we found that the performance of the HSS, D-Wave 2000Q QPU, and Advantage QPU show the potential of quantum annealing to solve industrial problems. Specifically, we note that the instance sizes tested using the QPUs are approaching the industrially-relevant limit. In the future, we will work on finding methods to mitigate the bottlenecks presented in this work, and other potential applications that lend themselves naturally to quantum optimization.  
\newpage
\bibliographystyle{unsrt}
\bibliography{references}






\end{document}